\documentclass[conference]{IEEEtran}
\IEEEoverridecommandlockouts
\usepackage{amsmath,amssymb, nccmath,amsthm}
\usepackage{cuted}

\newif\ifdraftmode
\ifdraftmode
    \usepackage{draftwatermark}
\fi
\usepackage{subcaption,graphicx}
\newcommand{\rulesep}{\unskip\ \vrule\ }
\usepackage{algorithmic}
\usepackage{graphicx}
\usepackage{enumitem}
\usepackage[backend=bibtex,giveninits=true,citestyle=ieee]{biblatex} 
\usepackage{textcomp}
\usepackage{hyperref}
\makeatletter
\g@addto@macro{\UrlBreaks}{\UrlOrds}
\makeatother
\usepackage{caption}
\usepackage{thmtools}
\newcommand{\RNum}[1]{\uppercase\expandafter{\romannumeral #1\relax}}
\declaretheoremstyle[
spaceabove=6pt, spacebelow=6pt,
headfont=\normalfont\bfseries,
notefont=\mdseries, notebraces={(}{)},
bodyfont=\normalfont,
postheadspace=0.6em,
headpunct=:
]{mystyle}

\usepackage{cleveref}
\crefname{hyp}{hypothesis}{hypotheses}
\Crefname{hyp}{Hypothesis}{Hypotheses}
\usepackage{glossaries}
\makeglossaries
\newacronym{isp}{ISP}{internet service provider}
\newacronym[plural={points of interest}, \glsshortpluralkey={POIs}]{poi}{POI}{point of interest}
\newacronym{od}{OD}{origin destination}
\newacronym{rog}{ROG}{radius of gyration}
\newacronym{gsm}{GSM}{Global System for Mobile Communication}
\newacronym{gps}{GPS}{Global Positioning System}
\newacronym{sdk}{SDK}{Software Development Kit}
\newacronym{cdr}{CDR}{Call Data Record}
\newacronym{xdr}{XDR}{X Data Record}
\newacronym{volte}{VoLTE}{Voice over LTE}

\usepackage{xcolor}
\usepackage{doi}

\DeclareMathOperator{\weightedMean}{weightedMean}
\DeclareMathOperator{\distanceHaversine}{distanceHaversine}
\DeclareMathOperator{\timeWeightedCentroid}{timeWeightedCentroid}
\DeclareMathOperator{\rog}{rog}
\usepackage{todonotes}
\AtEveryBibitem{%
    {\clearfield{archivePrefix}
    \clearfield{arxivId}
    }
    {}%
}
\DeclareSourcemap{
  \maps{
    \map{
      \step[fieldset=eprint, null]
      \step[fieldset=eprinttype, null]
      \step[fieldset=arxivId, null]
      \step[fieldset=archivePrefix, null]
      \step[fieldset=issn, null]
      \step[fieldset=isbn, null]
    }
    \map{
      \pernottype{misc} 
      \step[fieldset=url, null] 
    }
  }
}
\usepackage[section]{placeins}

\begin{document}

\title{
Country-wide mobility changes observed using mobile phone data during COVID-19 pandemic
}

\author{\IEEEauthorblockN{Georg Heiler\IEEEauthorrefmark{1}\IEEEauthorrefmark{2}, 
Tobias Reisch\IEEEauthorrefmark{2}\IEEEauthorrefmark{3}, 
Jan Hurt\IEEEauthorrefmark{2}, 
Mohammad Forghani\IEEEauthorrefmark{4}, 
Aida Omani\IEEEauthorrefmark{4}, 
Allan Hanbury\IEEEauthorrefmark{1}\IEEEauthorrefmark{2},
Farid Karimipour\IEEEauthorrefmark{5}
}
\IEEEauthorblockA{\IEEEauthorrefmark{1}Institute of Information Systems Engineering, TU Wien, Favoritenstr. 9-11, 1040 Vienna, Austria}
\IEEEauthorblockA{\IEEEauthorrefmark{2}Complexity Science Hub, Josefstädter Str. 39, 1080 Vienna, Austria}
\IEEEauthorblockA{\IEEEauthorrefmark{3}Institute for Complex Systems, Medical University of Vienna, Spitalgasse 23, 1090 Vienna, Austria}
\IEEEauthorblockA{\IEEEauthorrefmark{4} School of Surveying and Geospatial Engineering, College of Engineering, North Kargar Ave., Teheran, Iran (Graduated)}
\IEEEauthorblockA{\IEEEauthorrefmark{5}Edelsbrunner Group, Institute of Science and Technology (IST), Am Campus 1, 3400 Klosterneuburg, Austria}
\thanks{
This work was funded by the Austrian Research Promotion Agency (FFG) under project 857136, the Austrian Science Fund FWF under project P29252, the WWTF under project COV 20-017 and COV20-035 and the Medizinisch-Wissenschaftlicher Fonds des Buergermeisters der Bundeshauptstadt Wien under project CoVid004.
Corresponding author: F. Karimipour (e-mail: farid.karimipour@ist.ac.at).
}
}


\maketitle
\begin{abstract}
In March 2020, the Austrian government introduced a widespread lock-down in response to the COVID-19 pandemic. 
Based on subjective impressions and anecdotal evidence, Austrian public and private life came to a sudden halt.
Here we assess the effect of the lock-down quantitatively for all regions in Austria and 
present an analysis of daily changes of human mobility throughout Austria using near-real-time anonymized mobile phone data.
We describe an efficient data aggregation pipeline and analyze the mobility by quantifying mobile-phone traffic at specific \glspl{poi}, analyzing individual trajectories and investigating the cluster structure of the origin-destination graph. 
We found a reduction of commuters at Viennese metro stations of over 80\% and the number of devices with a radius of gyration of less than 500 m almost doubled.
The results of studying crowd-movement behavior highlight considerable changes in the structure of mobility networks, revealed by a higher modularity and an increase from 12 to 20 detected communities.
We demonstrate the relevance of mobility data for epidemiological studies by showing a significant correlation of the outflow from the town of Ischgl (an early COVID-19 hotspot) and the reported COVID-19 cases with an 8-day time lag. 	
This research indicates that mobile phone usage data permits the moment-by-moment quantification of mobility behavior for a whole country. We emphasize the need to improve the availability of such data in anonymized form to empower rapid response to combat COVID-19 and future pandemics.  
\end{abstract}


\section{Introduction}
\ifdraftmode
    \textbf{
    Social distancing is important, but how to quantify?
    }
\fi
It is generally agreed upon that ensuring a minimum spatial distance between people \cite{Greenstone2020} and limited exchange between segregated communities \cite{Abel2020,thurner2020most} are key factors in preventing the spread of COVID-19. After a surge in COVID-19 infections at the beginning of March 2020, the Austrian government introduced a widespread lock-down, asking its citizens to reduce their mobility and social contacts \cite{Desvars-larrive2020}. The call of the government seemed to be successful based on anecdotal evidence, such as reports of empty public spaces \cite{orf2020leerestadt} or low traffic levels on highways \cite{tt2020verkehr}. However, to estimate the effect on epidemic spreading and to plan further policy measures, a countrywide quantification of the effect of the measures was necessary.

\ifdraftmode
    \textbf{
    Use telephony data!
    }
\fi
Mobility information obtained from sources such as the \gls{gsm} network can be useful to monitor the reduction in mobility on a large scale \cite{Oliver2020}.
We applied this proposition to Austria and monitored daily changes of mobility in near-real-time using anonymized mobile phone data, compared behavior before, during and after lock-down measures and 
published parts of our results online 
\footnote{\url{https://csh.ac.at/covid19}}
due to the immanent relevance for the public.
Here we present and extend the results and elaborate on the technological background of our efforts during the COVID-19 pandemic.

\ifdraftmode
    \textbf{
    what we did 
    }
\fi
The analysis of mobility was performed for multiple spatial resolutions from points of interest to Austria as a whole, see Section \ref{s:resolution} for details.
Statistical measures such as the amount of travels, geometric measures such as \gls{rog} and activity space and graph-based approaches like local and global clustering coefficients were employed.
We observed a reduction of the usage of public transportation in the city of Vienna as well as an overall reduction of mobility and notable variation in the human mobility interaction network across Austria, indicated by the above-mentioned measures for the whole country.
In Section \ref{h:xischgl} we show analysis results connecting the spread of the disease with travels from the town of Ischgl, an early hot-spot in Austria.

We make the following contributions:
\begin{itemize}
    \item Design and implementation of an efficient processing pipeline that prepares mobile phone data on a daily basis for anonymized and aggregated mobility analyses.
    \item Demonstration of the influence of the lock-down in Austria on the values of the following metrics: \gls{poi} based counting, \gls{rog}, activity-space, graph-based community analysis. To our best knowledge, this is the first time that all these metrics are calculated from the same dataset.
    \item Confirmation of the usefulness of mobile phone data for modelling disease spread by identifying a significant correlation with a time-shift of 8 days for the outflow of people from the highly infectious quarantined region Ischgl in Austria to other municipalities.
\end{itemize}

\section{Related studies}
Extensive literature using mobility data during the COVID-19 pandemic has been published \cite{Pepe2020, JaysonS2020, Gao2020, Jeffrey2020, Yabe2020,Vollmer2020,Xu2020, JRC2020b, JRC2020, JRC2020a, Heuzroth}.
International mobility reports of mobile-phone based data are analyzed by the European commission to report on the effect of the COVID-19 lock-down including the comparison of the effect in different countries and estimation of cross-border effects \cite{JRC2020b, JRC2020, JRC2020a}.
Furthermore, \cite{Heuzroth} published analyses of mobility behavior subsequent to the lock-down using \gls{gps}-data evaluating the spread of the virus from the highly infectious region in Ischgl to different countries. 
They used data from the private company Umlaut, which collected \gls{gps} measurements using tracking toolkits in apps.
This data offers a higher positioning accuracy, however, the number of users and, hence, the number of data records and the coverage of the population are limited when compared with \gls{gsm}-based data.

Many studies, mostly employing data from China, analyze mobility with the objective to predict infection numbers.
Jia et al. \cite{JaysonS2020} and Goa et al. \cite{Gao2020} predict the number of infections in China with the outflow of people from Wuhan.
Kraemer et al. \cite{Kraemer2020}, Jeffrey et al. \cite{Jeffrey2020} and  Yabe et al. \cite{Yabe2020} report that the correlation between mobility and the infection rates dropped after implementation of the lock-down measures in China, the United Kingdom and Japan, respectively.

The two companies behind Android and iOS, Google\footnote{\url{https://www.google.com/covid19/mobility/}} and Apple\footnote{\url{https://www.apple.com/covid19/mobility}}, both published aggregate mobility reports as well.
These are available for many countries.
Vollmer et al. \cite{Vollmer2020} and Xu et al. \cite{Xu2020}
use mobility data to calibrate epidemiological models.
We extend the literature by investigating Austrian mobility behavior and combining a wide variety of measures.

\section{Materials and methods}
\subsection{Spatial resolution}
The analyses were performed at a multitude of spatial resolutions, on one hand to match the resolution of complementary data sources and on the other hand to address the resolution requirements of various research questions.
\label{s:resolution}
We use the following levels, which are increasing in the level of detail: Austria as a whole, federal states, political areas\footnote{\url{https://www.statistik.at/web_de/klassifikationen/regionale_gliederungen/politische_bezirke/index.html}}, post codes, municipalities and specific \glspl{poi}.

\subsection{Datasets}
\label{sec:datasets}
\subsubsection{Mobile phone usage data}
\label{sec:mobile_phone_usage_data}
Anonymized data was provided by a large Austrian \gls{isp}.
The information on all exchanges made between a mobile phone network and its users are recorded as \emph{events}.
Any direct (user plane) as well as indirect (control plane) interaction with the network continuously generates events in the dataset, which are aggregated daily.
The dataset is based on classical \gls{cdr} and includes \gls{xdr} of the data domain, thereby providing anonymized metadata about voice and data usage.
The events stem from various network interfaces covering all most widely used signalling technologies (2G, 3G, 4G, calls, text messages \& \gls{volte}).

The dataset contains approximately 1 Billion events from 4.5 Million devices.
For 80\% of these, the subsequent event is received in 1.7 minutes, on average 4 minutes.
This means that for some old, i.e. 2G devices, which are only rarely used, almost no data is transmitted whilst not actively in operation.
Therefore, fewer events are generated and these devices are thus much harder to analyze when considering them for mobility use cases.
Our analyses are based on filtering the data of approximately 1.2 Million devices registered with the partner \gls{isp} as mobile handsets excluding sensor devices from the Internet of Things as well as roamers\footnote{Devices with a foreign SIM-card using the local network, i.e. mostly tourists.} or events obtained from virtual network operators\footnote{Virtual operators resell the existing network of the providers -- often more cheaply.}.

The \gls{gsm} network registers events for each device $i$ with a very accurate time information and a location with latitude ($y_{lat}$) and longitude ($x_{long}$).
As the network continuously generates events, a near-real-time monitoring of aggregate population behavior is possible \cite{Xu2020}.
Our analyses were updated on a daily basis.

\emph{Localization:}
\label{s:localization}
Each cell-id of the network topology is localized with a $(x_{long}, y_{lat})$ tuple.
The localization information is provided by the \gls{isp}.
The accuracy of cell-id based localization is by far worse than \gls{gps}. 
However, \gls{gsm} data is more readily available and covers larger sample sizes.

\emph{Privacy:}
\label{sec:privacy}
Obeying data privacy regulations is important when analyzing mobile phone usage data.
We analyze only k-anonymized aggregated data where tracking of individuals for extended periods of time is impossible.
Due to data privacy restrictions, any identifiers (primary keys identifying a subscriber) are anonymized.
This is handled by encrypting them with a rotating key, which is changed every 24 hours by the \gls{isp}.
This means that analyzing identifiers over more than 24 hours is impossible.
Additionally, we use only cell-id based localization to enhance the privacy of the subscribers due to its limited accuracy.
Thereby the local regulations have been met and the recommendations of the GSMA, the alliance of mobile phone providers \cite{gsmacovid} have been followed.

\subsubsection{Infection dataset}
\label{data:ages}
On behalf of the Austrian Ministry of Health, the company \emph{Gesundheit Österreich GmbH} provides access to the electronic epidemiological reporting system (EMS)\footnote{\url{https://datenplattform-covid.goeg.at/}}.
The data contains cumulative daily COVID-19 infection numbers from March 5\textsuperscript{th} onwards.

\subsection{Scalable data aggregation pipeline}
\begin{figure*}
  \includegraphics[width=\textwidth]{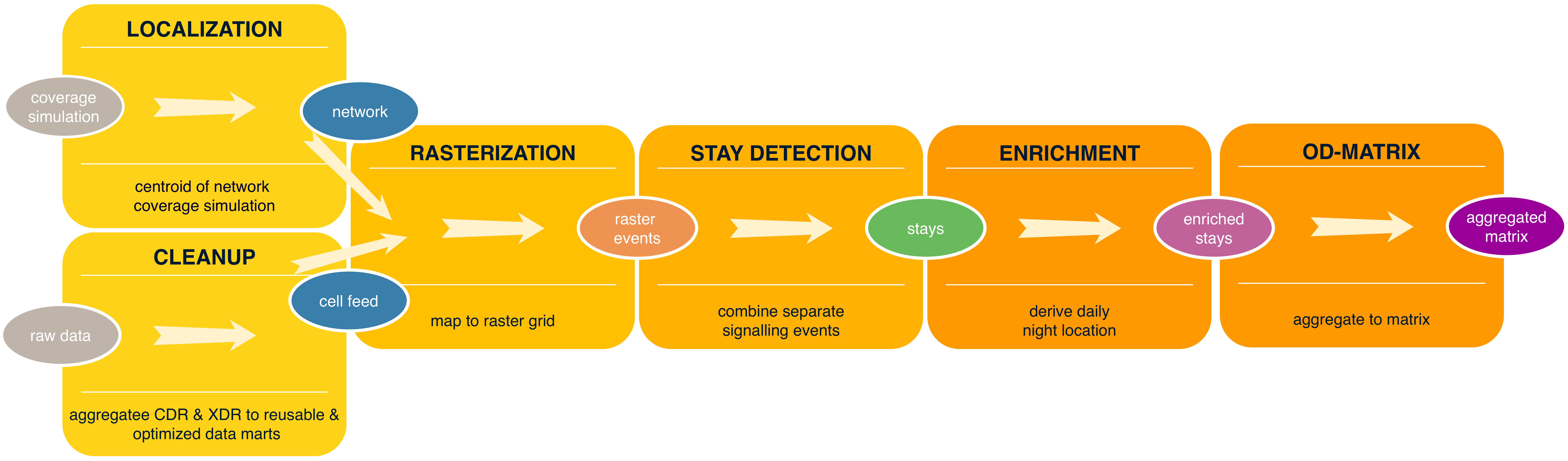}
  \caption{Daily aggregations are calculated as a first step to clean up and compress the data. Subsequent analyses can be implemented efficiently -- as depicted here in case of the computation for an \gls{od} matrix.}
  \label{fig:DataFlow}
\end{figure*}

Our data processing pipeline is depicted in Figure \ref{fig:DataFlow}.
Firstly, to improve the performance of our analyses and to ease the mental burden for the person conducting the evaluation, we computed a daily aggregation of the raw data, removing domain-specific knowledge, prior to beginning the specific analyses.
The \gls{gsm} network and attached monitoring tools are a very complex system which requires a lot of business knowledge (telecommunication specific knowledge required to process the raw data, such as special types of events or structures of the topology).
We abstracted this knowledge away and allow for effective and efficient analyses on top of our aggregations.

For each device, timestamp and cell-id were recorded and anonymized with a rotating key. Then, we enriched each cell-id with its location information which is provided by the \gls{isp} (described in \ref{sec:mobile_phone_usage_data}).
As a next step, stays within a raster cell, which are defined by the regional levels listed in Section \ref{s:resolution}, were detected by spatio-temporal clustering the signalling events.
Thereafter, each stay was enriched with the daily night location, see Section~\ref{sec:homelocation} for details.
Finally, the \gls{od} matrix was created as described in Section \ref{sec:od}.
After the daily cleanup, aggregation and compression, analyses were built on this solid and reusable foundation.

We needed to process an immense quantity of data as the raw events amount to more than one billion per day.
Therefore, we rely on a cluster of computers to achieve good performance.
Using Apache Spark \cite{Zaharia2012}, a map-reduce style big-data framework, the burden of a distributed system is eased for the developers of the analyses as failures of compute nodes are handled automatically.

As Spark lacks support for geospatial primitives, 
GeoSpark was utilized for distributed spatial joins  \cite{Yu2015}.
Additional geospatial functionality is made available by using GeoMesa \cite{Hughes2015},
which in its core is based on the java-topology-suite for providing the geospatial functionality.
This is similar to GeoSpark, but still provides complimentary functionality such as calculation of spatial distances.
For the enrichment of spatial metadata such as political areas, we utilize a custom broadcast spatial join \cite{Heiler2019}, which is faster than its distributed equivalent.

Furthermore, columnar file formats with compression and run-length encoding are utilized, as these allow for significant compression of the data when sorted by the join keys.

\subsection{Quantification of mobility}
We quantify the movement of the population by several measures ranging from simple counting to estimation of the mobility via \gls{rog} and activity space, to the evaluation of \gls{od}-flow matrix, which are described in the following subsections.

\subsubsection{Counting devices}
The mobile phone network consists of multiple topological layers.
Some base stations cover a larger area.
Indoor or underground cells as for example found in metro stations are well suited to monitor the number of daily commuters due to their small coverage area.
For several \gls{poi}, i.e. base station in the Viennese underground metro network or for an airport and region of quarantine (Ischgl) we count the number of unique subscribers per day.

\subsubsection{Night location}
\label{sec:homelocation}
Based on the proposed algorithm of \cite{Widhalm2015} we utilize the most used cell towers to derive a home location.
We only use night time activity from 8 pm until 12 pm to obtain a spatial reference for each device $i$.
The resulting spatial reference is mapped to a post-code for further analysis.

\subsubsection{\glsfirst{rog}}
\label{sec:rog}
The \gls{rog} is defined as the time-weighted root mean square distance between the center of gravity and each individual localization, see Equation \ref{eq:rogtw}.
The center of gravity is calculated as the coordinate-wise time-weighted average, as defined in Equation~\ref{eq:centroid}.
We use the Haversine distance to obtain the distance in Meters, cf. Equation \ref{eq:haversine}.
\begin{table*}
\captionsetup{labelformat=empty}
\caption{Calculation of \gls{rog}}
\begin{minipage}{\textwidth}
\begin{align} 
  \weightedMean(\vec{x},\vec{w}) &= \frac{\sum_{i}^{n} w_i * x_i}{\sum_{i}^{n} w_i} \label{eq:weightedMean} \\
  \timeWeightedCentroid(\vec{x_{long}},\vec{y_{lat}},\vec{w}) &= \left[ \weightedMean(x_{long}, w), \weightedMean(y_{lat}, w) \right] \label{eq:centroid} \\
  \distanceHaversine(\vec{x_{long_1}},\vec{y_{lat_1}},\vec{x_{long_2}},\vec{y_{lat_2}}) &= 2r \arcsin{\sqrt{\sin^2{\left(\frac{\vec{y_{lat_1}}-\vec{y_{lat_2}}}{2} \right)}+ \cos(\vec{y_{lat_1}}) \cos{(\vec{y_{lat_2}})} \sin^2{\left(\frac{\vec{x_{long_1}}-\vec{x_{long_2}}}{2} \right)}}} \label{eq:haversine}\\
  \rog(\vec{x_{long}},\vec{y_{lat}},\vec{w}) &= \sqrt{\weightedMean \bigl( \distanceHaversine(\timeWeightedCentroid(\vec{x_{long}}, \vec{y_{lat}}, \vec{w}), \vec{x_{long}}, \vec{y_{lat}})_i^2, \vec{w} \bigr)} \label{eq:rogtw}
\end{align}
\medskip
\hrule
\end{minipage}
\end{table*}

\subsubsection{OD flow matrix construction}
\label{sec:od}
The mobility interaction network can be captured by extracting the origin–destination (\gls{od}) matrix, which specifies the amount of travel between regions throughout the study area.
It is calculated for multiple scales including macroscopic scales, e.g., at the inter-urban level, or at microscopic scales, e.g., at the intra-urban level.
In recent years, OD matrices have often been constructed from mobile phone data \cite{Sohn2008,Mamei2019,Guo2018,Purnama2018,Louail2015,Bachir2019,FORGHANI2020102666}.

A \emph{trajectory} can be modeled by sorting the localized events per user by time.
To derive the \gls{od} matrix, the continuous stream of point localizations in the network is first rasterized to the desired resolution.
We are analyzing various resolutions as defined in Section \ref{s:resolution}, e.g. municipalities and post-codes as well as mathematically well defined grids like Uber's H3 \cite{uber_h3}.
For each discrete location $l$ a stay duration is computed, which is referred to as weight $w$.

We cluster these discretized point localizations by space and time in order to compute time-weighted stays for each user and raster cell.
The most important points can be aggregated as a \gls{od} matrix, where \emph{most important} refers to the points with a stay duration of at least $s_k$ seconds.
Each stay has an associated entry and exit time.
We set a threshold of $s_k=600$ seconds for our analyses.
Finally, we aggregate the matrix over all the devices $i$ daily by counting the subscribers moving from one grid unit to the other.

\subsubsection{Activity space}
The activity space of a certain discrete location is the environment or area in which its population moves.
In other words, it indicates the regionalized dispersion of the places that people visit.
Among different forms of activity space representation, here we use the ellipsoidal representation fitted to all the visited points \cite{10.1007/978-3-319-95165-2_33}.
Compared to the \gls{rog} (Sec. \ref{sec:rog}), an ellipse provides a more accurate representation of the locations by computing the major and minor axes as well as an angle.
We use the \gls{od} matrix to compute the following parameters for the activity space ellipse of each political area \cite{10.1007/978-3-319-95165-2_33}:
\begin{itemize}
    \item Length of major and minor axes
    \item Area
    \item Azimuth: The angle between the north direction and the major axis
    \item Shape index: An indication of how much the shape of ellipse is close to a circle (which is 1.00 for a circle)
\end{itemize}
Moreover, the amount of intra- and inter-region travels as well as the sum of distance of travels and average distance of travels can be extracted from the \gls{od} matrix. 
For geometric investigation of the mobility behavior at the federal state level, the above parameters were averaged over the political areas of each state.

\subsubsection{Graph-based movement analysis}
The use of graph-based analyses in crowd-movement studies has been investigated, especially in the use of mobility data extracted from cellular networks \cite{ghahramani2020urban}. 
The \gls{od} matrix can be interpreted as a graph where pairs of nodes $m$ and $n$ represent
origins and destinations which are connected by links with non-negative weights $A_{m,n}$ if one or more trips are made between the nodes.
By modeling the crowd-movement in the structure of a graph, it is possible to characterize the architecture and dynamics of the population mobility and demonstrate relationships between people and places
\cite{forghani2018interplay}.
In graph theory, the topological criteria such as centrality, connectedness, path length, diameter, and degree play a key role in the description of a graph where links are usually represented as binary states (i.e. adjacency matrix). For the analysis of mobility, the difference in strength of the interaction links between pairs of nodes is important. \cite{saberi2017complex}.

We first utilize the \emph{Local Clustering Index} and \emph{Global Clustering Index} to more deeply understand the movement-based interdependencies and interactions between locations and then perform community detection to compare the structure of the network during multiple time periods.
   
\emph{Local clustering coefficient:}
This metric 
measures the degree to which the nodes in a graph tend to cluster together and is defined for any node $m$ as the fraction of  total relative weight of connected neighbors of that node over the total mobility strength of possible links between the node's neighbors. This measure is calculated by \cite{barrat2004architecture}:
\begin{equation}
    LC_m = \frac{1}{s_m (k_m -1)} \sum_{n,h}\frac{A_{m,n} + A_{m,h}}{2} a_{m,n}a_{m,h}a_{n,h}
    \label{eq:local_clustering}
\end{equation}
where $A$ represents the weight of each edge between a pair of nodes, $s_m$  is sum of weights of edges connected to node $m$, $k_m$ is the degree of node $m$ and $a$ is the binary state of presence of a link (0 is assigned for each \gls{od} pair that does not have an observed flow and 1 is assigned otherwise).
Generally, well interconnected nodes show a high local clustering coefficient while the nodes with neighboring vertices that are not directly connected present a small clustering coefficient.

\emph{Global clustering coefficient:}
This metric is based on transitivity, which is calculated based on the weighted density of triplets of nodes in a graph.
A triplet can be defined as three nodes that are connected by either two (open triplet) or three (closed triplet) links. The global clustering coefficient is defined as the number of closed triplets over the total number of triplets \cite{opsahl2009clustering}:
\begin{equation}
    GC=\frac{\sum_{\tau \Delta} w
    }{\sum_{\tau} w
    }
    \label{eq:global_clustering}
\end{equation}
where $w$ is defined as the arithmetic mean of the weights, 
$\sum_{\tau} w$ is the total weight of triplets
and  $\sum_{\tau \Delta} w$ is total weight of the subset of the closed triplets.

\emph{Community detection:}
The mobility interaction network described above is neither regular like lattices nor irregular like a random graph with fairly homogeneous distribution of edges among the nodes.
The movement network consists of a heterogeneous distribution of edges both globally and locally.
A community in such a network might refer to a group of nodes presenting a large number of internal links as well as a small number of links toward the nodes pertaining to other communities. The notion of \emph{modularity} is the most broadly used measurement of the quality of a network division into communities. For a weighted graph, the modularity is calculated by \cite{newman2006modularity}:
\begin{equation}
    Q=\frac{1}{2M}\sum_{m,n}(A_{m,n} - \frac{k_m k_n}{2M} \phi(c_m, c_n))
    \label{eq:community_detection}
\end{equation}
 where $M$ defines the total number of edges, $A_{m,n}$ represents the edge weight between nodes $m$ and $n$, $k_m$ and $k_n$ indicate the degree of the vertices, $c_m$ and $c_n$ are the communities of  $m$ and $n$, and $\phi(c_m, c_n)$ equals 1 when $m=n$ and 0 otherwise. 
 If a particular division gives no more within-community edges than a randomly divided network we will get $Q=0$. Values other than 0 indicate deviations from
randomness, and in practice values greater than about 0.3 appear to indicate significant community structure. Given that high values of modularity represent good partitions, the best partitioning or at least a very good partitioning of a graph could be attained through optimization methods.
Newman \cite{newman2004fast} introduces a well-known community detection algorithm. In this algorithm, starting with a state in which each node is the sole member of
one individual community, a greedy algorithm is employed for repeatedly joining communities together in pairs and choosing, at each step, the join that results in the greatest increase in $Q$. This process continues until the corresponding maximal modularity is achieved.

\section{Results}
In our analyses we consider multiple phases indicated by Roman numerals, based on the dataset of non-medical interventions published by Desvars et. al. \cite{Desvars-larrive2020}: 
\begin{enumerate}[label=\Roman*.]
\item pre-lock-down -- before 11\textsuperscript{th} of March
\item transitioning -- 12\textsuperscript{th} until 14\textsuperscript{th} of March
\item lock-down -- 15\textsuperscript{th} of March until 1\textsuperscript{st} of May
\item easing -- 2\textsuperscript{nd} of May onwards
\end{enumerate}

\subsection{Counting-based evaluation}
\subsubsection{Public transport usage}
\begin{figure}
\centering
  \includegraphics[width=\columnwidth]{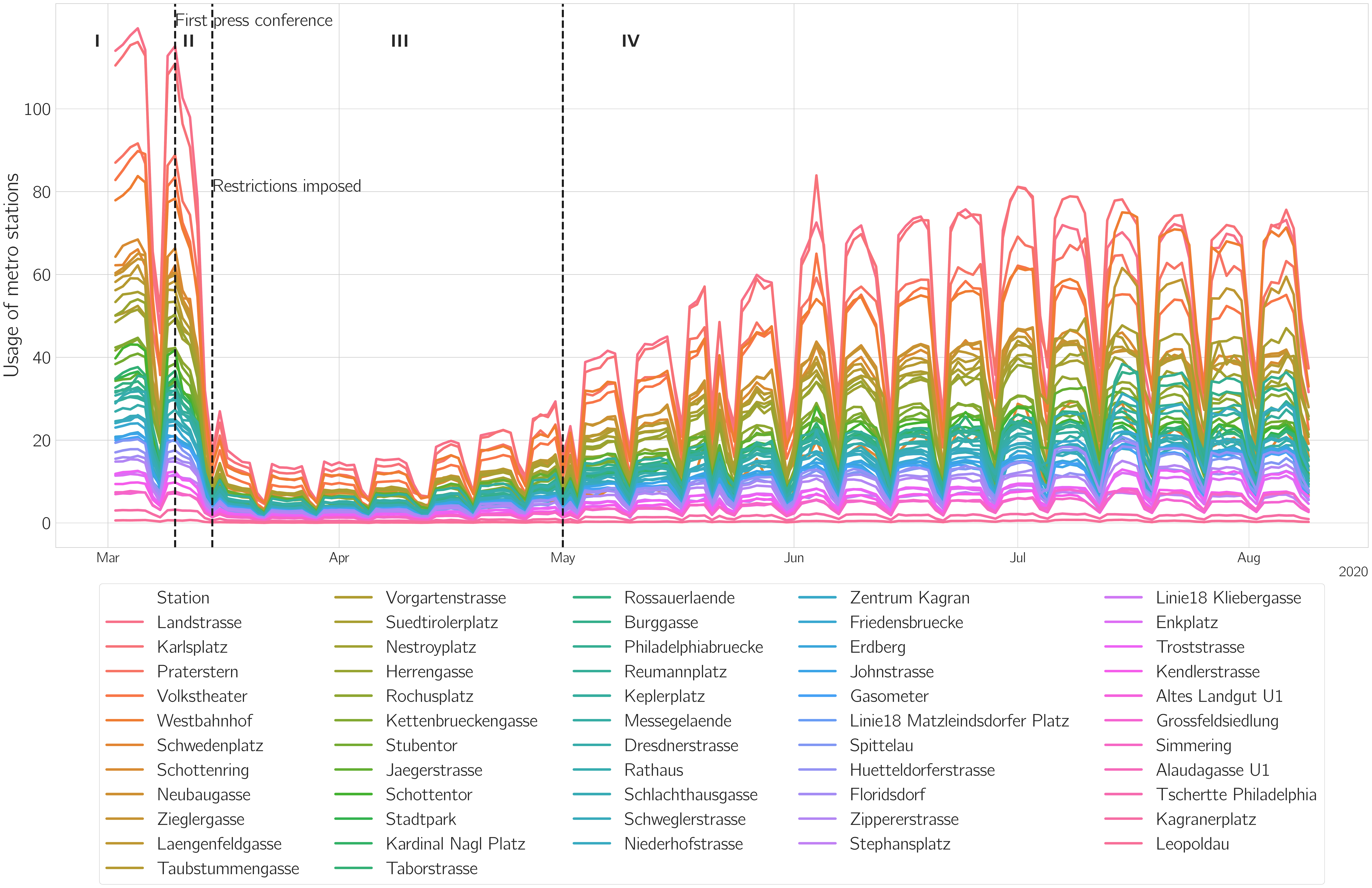}
  \caption{Reduction in public transport usage in Vienna during the COVID19 pandemic.
  Even after easing of the measures previous levels of metro usage are not reached again.}
  \label{fig:metro}
\end{figure}

Figure \ref{fig:metro} shows the reduction of passengers on the Viennese metro, which translates into the effectiveness of the far-reaching restrictions undertaken by the government of Austria.
After a first press conference on the 10\textsuperscript{th} of March (first black line), the measures were announced and activity was reduced until full implementation of the lock-down measures on the 15\textsuperscript{th} of March.
The frequency of metro usage was about $1/5$ of a regular Monday in this state induced by full implementation of the lock-down measures. 
We still can observe the weekly trend that there is less usage of the metro during weekends.
From Easter onwards
metro usage starts to recover almost to previous levels.
With the official end of the lock-down, mobility has recovered to 52.5\% when comparing calendar week 22 with week 10 -- i.e. with the levels of before the crisis.
Even later until August a full recovery to previous levels is not reached.

\subsubsection{\gls{poi} analyses}
\begin{figure}
  \includegraphics[width=\linewidth]{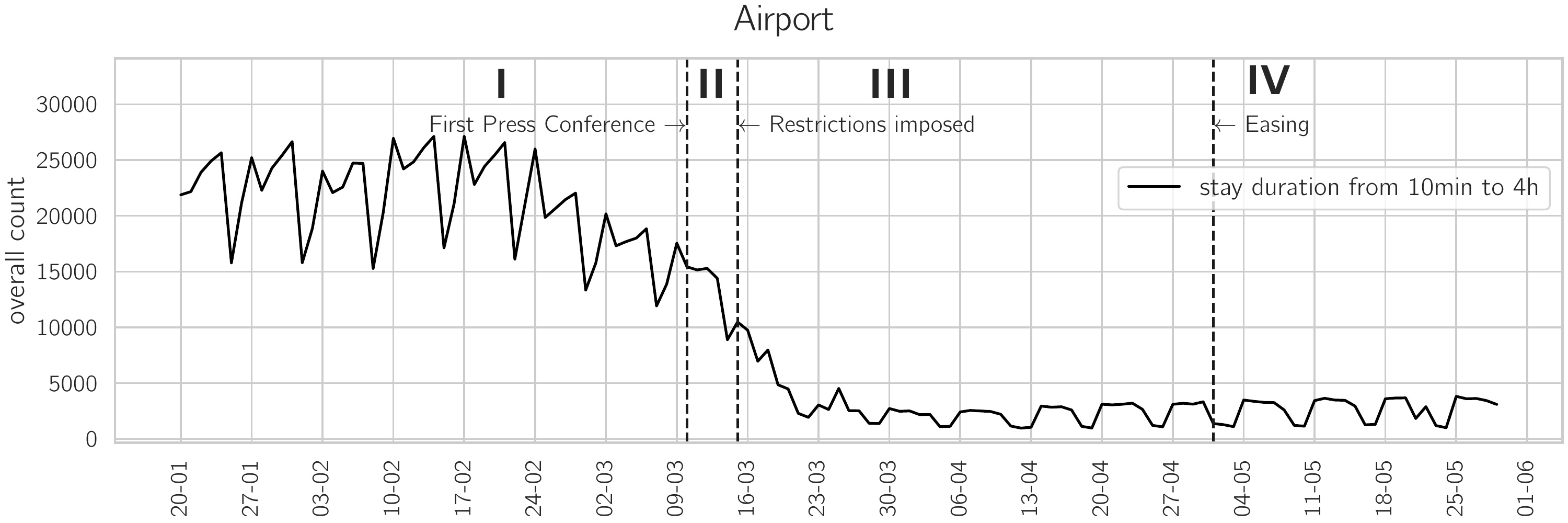}
  \includegraphics[width=\linewidth]{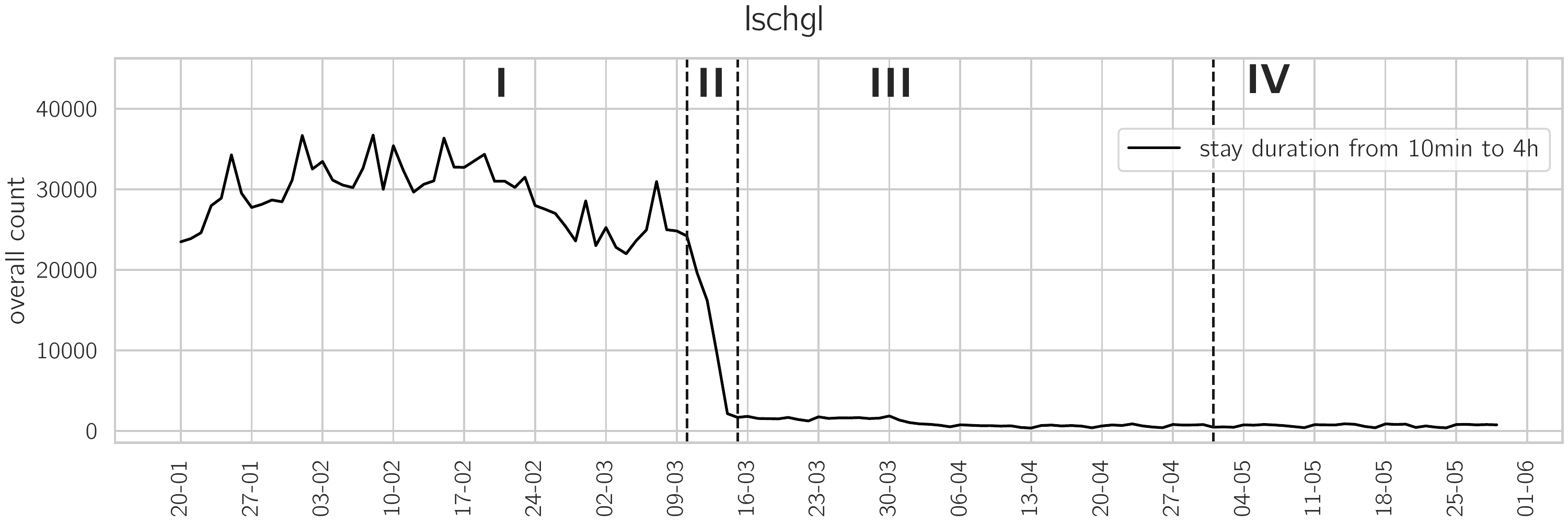}
  \caption{The 
  count of mobile phones with a stay duration from 10 minutes to 4 hours for Ischgl and the Airport of Vienna.
  There is a clear difference between the Airport and Ischgl.
  While Ischgl went into quarantine, and all tourists were sent home on the 15\textsuperscript{th} of March, the shutdown of the Airport happened the following week.}
  \label{fig:pis}
\end{figure}

For two selected locations (airport, quarantined region), see Figure \ref{fig:pis}, the dramatic reduction in devices present is depicted.
Both can be seen as a proxy for long distance/international travel activities\footnote{\url{https://www.tirolwerbung.at/wp-content/uploads/2018/04/tiroler-tourismus-daten-und-fakten-2017.pdf}}.
This also justifies that both locations have not recovered until the end of the analysis period.

\subsection{\gls{rog}}
\begin{figure}
\centerline{\includegraphics[width=\linewidth ]{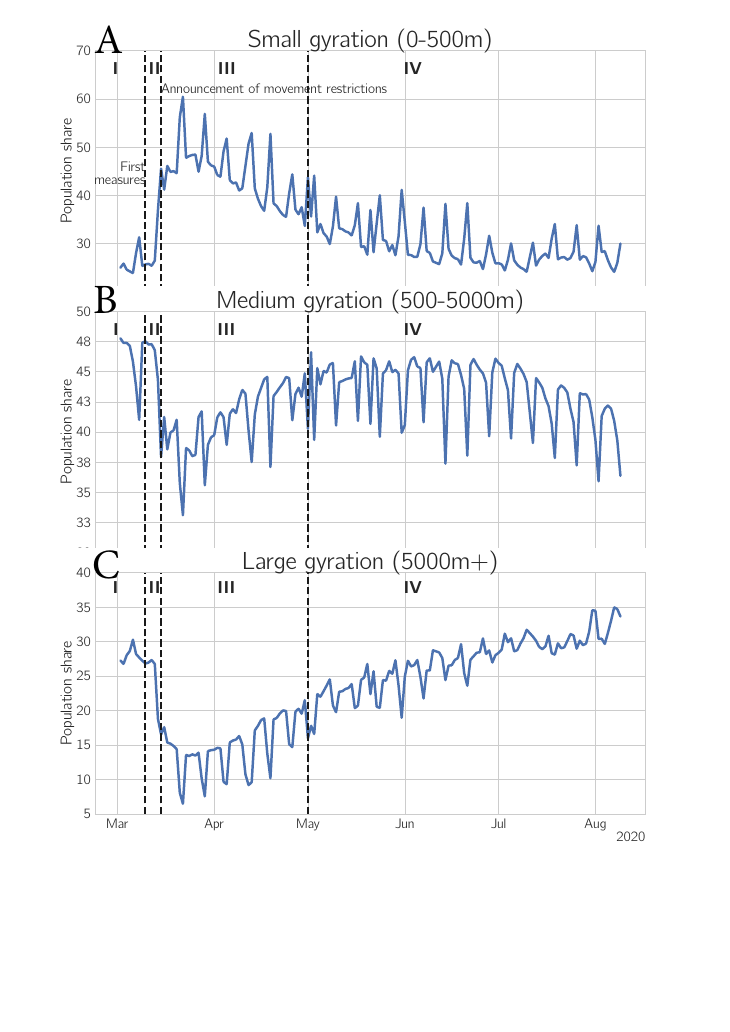}}
\caption{Bucketed \gls{rog} into small \textbf{A} (0 m - 500 m), medium \textbf{B} (500 m - 5000 m), large \textbf{C} ($>$5000 m) movement.}
\label{fig:ROG_medium}
\end{figure}

Before the crisis, the median \gls{rog} for the whole population was 2 kilometers per day. After the announcement of the restrictions on the 15\textsuperscript{th} of March, it reduced to 800 meters.
The distribution of the \gls{rog} is heavily skewed.
When creating discrete bins of the \gls{rog} the effect of very large \gls{rog} can be mitigated. 
We create three bins, $[0,500\mathrm{m}[$ for devices moving little to no movement, $[500,5000\mathrm{m}[$ for intermediate and $[5000\mathrm{m}, max.]$ for large movements. Bin sizes were chosen based on qualitative experience with test devices. Figure~\ref{fig:ROG_medium}A depicts how the lock-down measures increased the number of devices moving very little.
Even after the official easing of the measures the population has not yet recovered to previous levels of movement until the end of our study period.
Conversely, for medium distance movements in the range of 500--5000 m and large radii above 5000 m, Figures \ref{fig:ROG_medium}BC show the effect of the lock-down measures by depicting a dramatic reduction of movement.

\begin{figure}
  \includegraphics[width=\columnwidth]{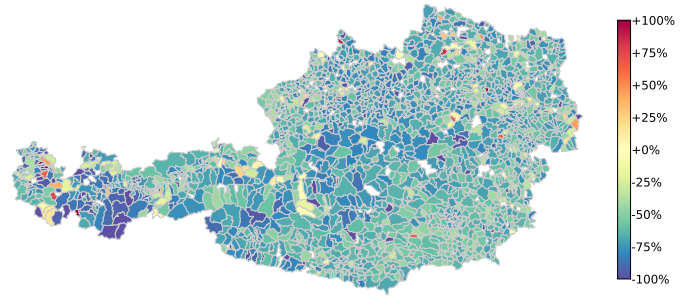}
  \caption{Relative change of mean \gls{rog} for week of March 2\textsuperscript{nd} and week of March 23\textsuperscript{rd} measured at postcode level.}
  \label{fig:rel_reduction_ROG_austria}
\end{figure}

We additionally computed a daily night location as defined in Section \ref{sec:homelocation} for each user.
This location was assigned the matching post code allowing to produce a map of Austria, Figure \ref{fig:rel_reduction_ROG_austria}, visualizing the spatial differences in the relative reduction of mobility, as measured by the \gls{rog}.
The reduction is consistent throughout Austria, except for some small towns where the number of observations might be too small.
\begin{figure}
\centerline{\includegraphics[width=\linewidth ]{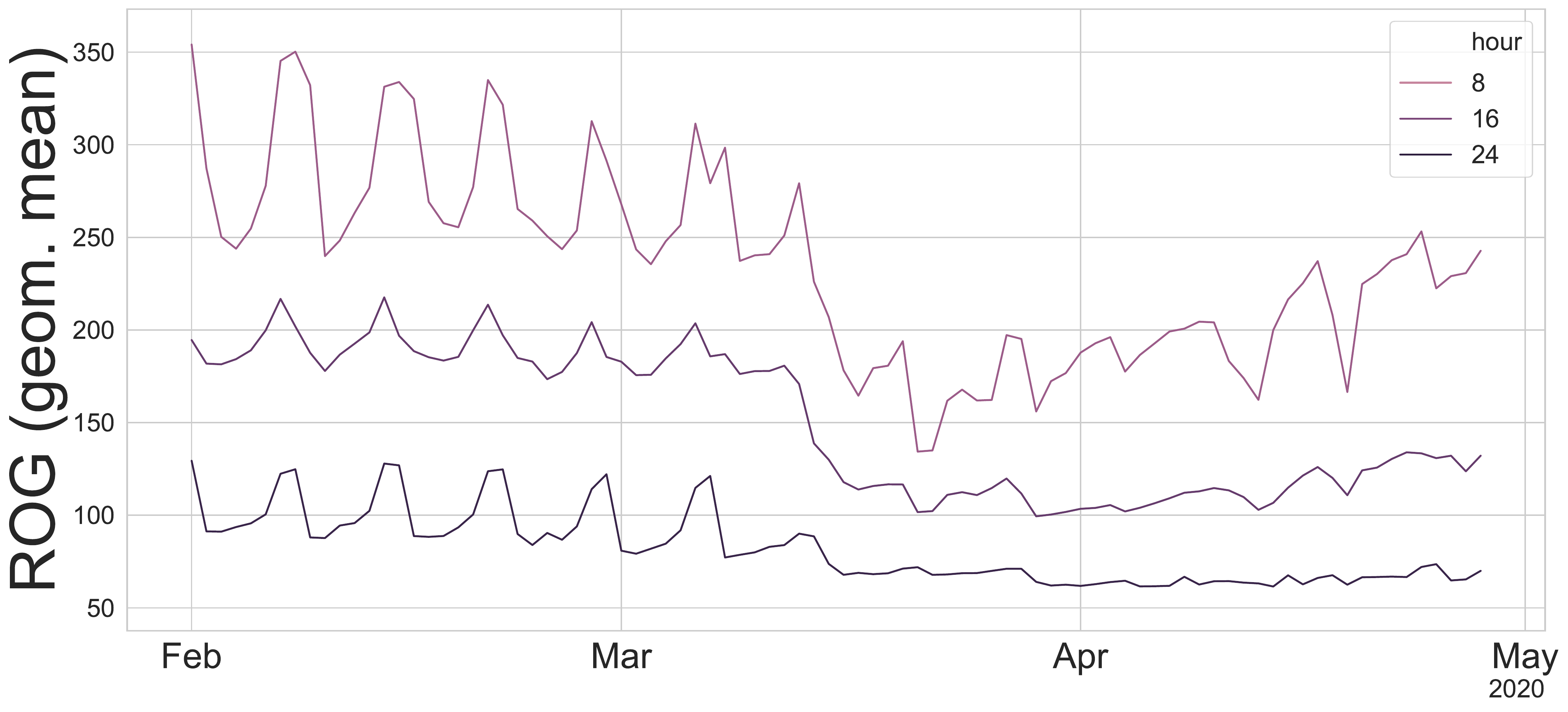}}
\caption{Hourly geometric mean of \gls{rog} for selected hours. The night activity on weekends is not recovering, even after reduction of the lock-down restrictions. }
\label{fig:rog_hourly}
\end{figure}
Moreover, we have broken down the analysis into hourly groups (Figure  \ref{fig:rog_hourly}).
Each line represents an hour. 
Before the lock-down, we observe a fairly consistent weekly trend. 
During the week there is a large spread between daily and nightly mobility, whereas on weekends this gap is reduced strongly due to less movement during the daytime and an increase in \gls{rog} at night.
With the introduction of the lock-down measures \gls{rog} decreases for all times, but retains its weekly pattern, except for the characteristic nightly increase of activity on weekends, which is not recovering, even after reduction of the lock-down restrictions.
Apart from weekend nights mobility at all times of day is slowly increasing towards pre-lock-down level.

\subsection{Activity space}
\begin{figure*}
  \includegraphics[width=\textwidth]{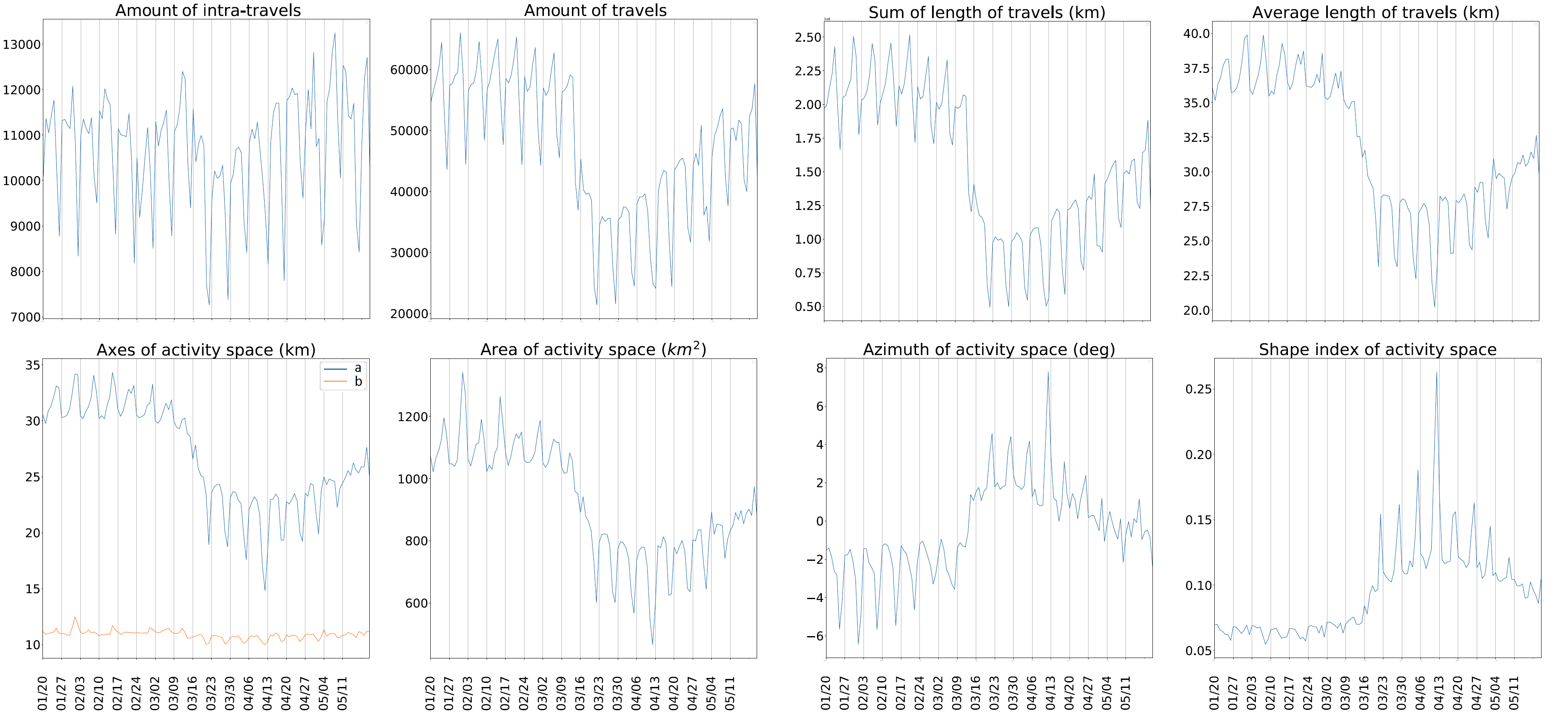}
  \caption{Absolute values of the activity space parameters for a sample political area.
  The results indicate a drastic reduction in mobility.}
  \label{fig:Activity_Spc_Abs}
\end{figure*}

The result of activity space analysis supports the above findings. Figure \ref{fig:Activity_Spc_Abs} illustrates a reduction in the statistical and geometrical parameters for an example political area from 1\textsuperscript{st} of March and especially the announcement of restrictions (on 15\textsuperscript{th} of March) until Easter (12\textsuperscript{th} of April). However, after Easter and especially after easing the restrictions (on 2\textsuperscript{nd} of May), the parameters are getting close to their normal values.
Specifically, as seen in Figure \ref{fig:Activity_Spc_Abs}, the length of the major axis severely reduces during the restriction time, while the minor axis is unchanged (this is illustrated in the graph of the shape index, too).
Moreover, as the graph of angle shows, the direction of the ellipse follows a periodic pattern during the restricted mobility time.
It means that destination points have become more specific, compared to randomly distributed destinations, which results in random values of angle before the restriction (note that when destination points are randomly distributed, the angle parameter varies from day to day, however, if the same destinations are visited every day, this parameter is almost similar for different days). This pattern has been followed by almost all political areas.
\begin{figure*}
\begin{subfigure}{\columnwidth}
    \includegraphics[width=\columnwidth]{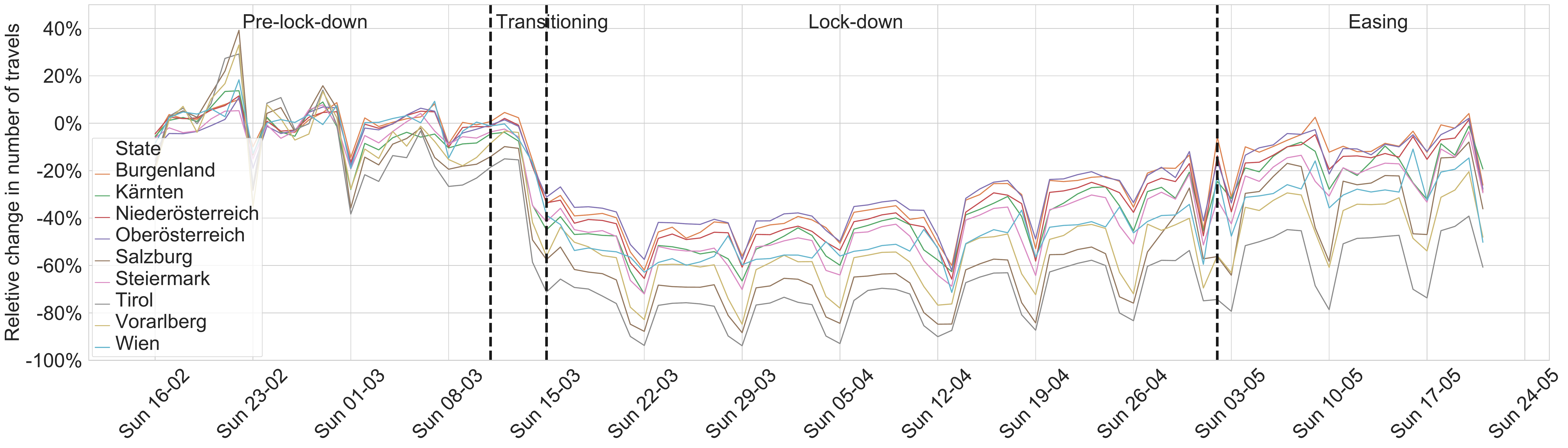}
    \includegraphics[width=\columnwidth]{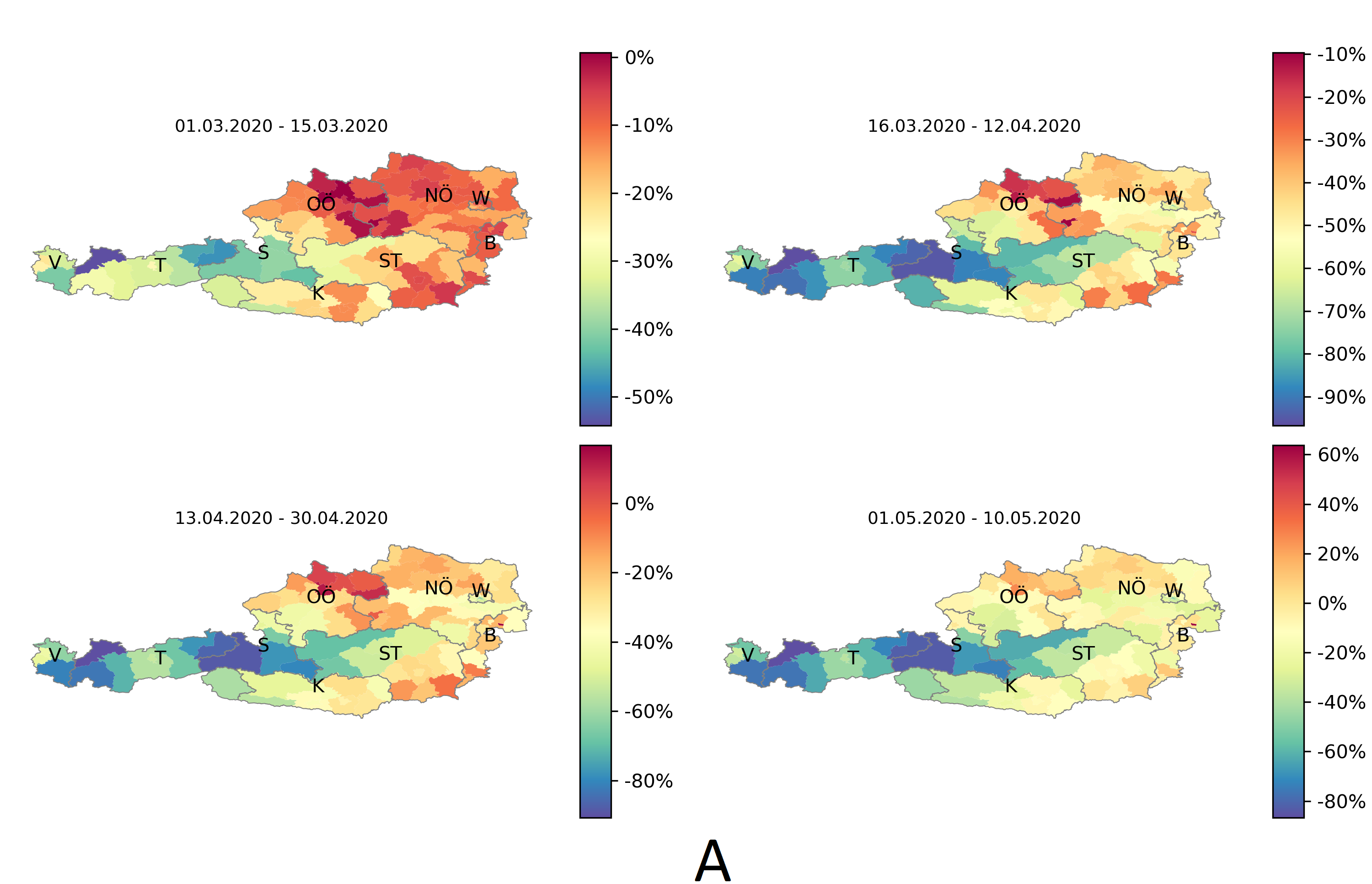}
\end{subfigure}
\rulesep
\begin{subfigure}{\columnwidth}
    \includegraphics[width=\columnwidth]{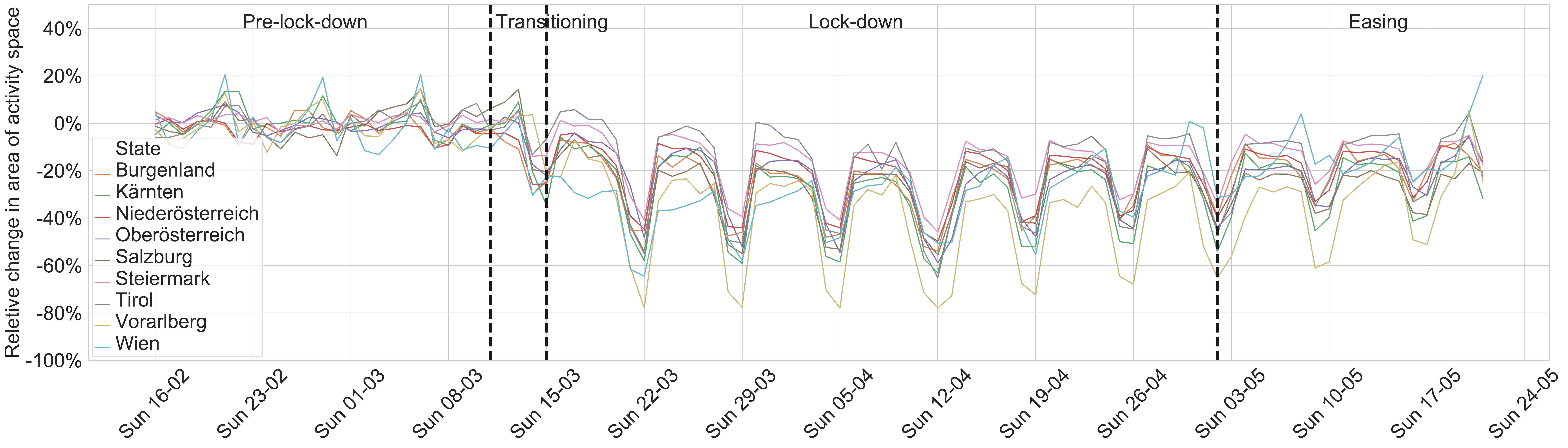}
  \includegraphics[width=\columnwidth]{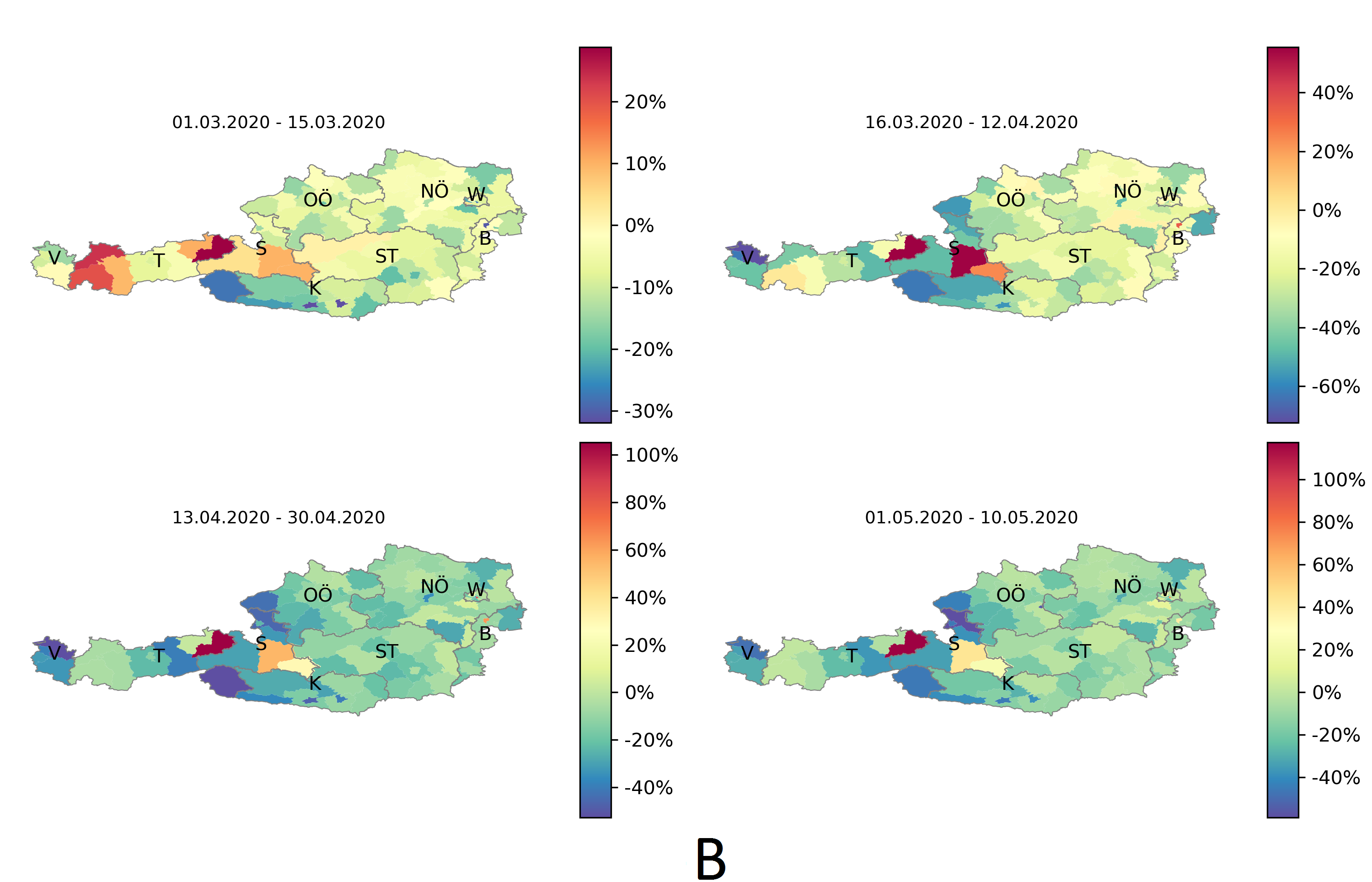}
\end{subfigure}
\caption{\textbf{A}: Relative change in sum of intra- and inter-region travel; \textbf{B}: Relative change in area of activity space. All metrics are broken down by Austrian federal states}
\label{fig:ellipses}
\end{figure*}
Figure \ref{fig:ellipses} (upper panels) plot both the relative change in the sum of intra- and inter-region travel as well as area of the activity space for federal states.
The corresponding parameters are shown on the maps in Figure \ref{fig:ellipses} (lower panels).
These figures illustrate a reduction in the relative change in the parameters for all states.
On the other hand, these figures show that the western states (i.e. Tyrol, Vorarlberg, and Salzburg), where the first positive COVID-19 cases were reported, started the reductions earlier than other states, even before official announcement of the restrictions, and have the highest reduction rates during the restriction periods. Moreover, they are still showing some levels of reduction although the restrictions have been eased on the 2\textsuperscript{nd} of May.

\subsection{Graph-based analysis of movement}
\begin{figure}
  \includegraphics[width=\columnwidth]{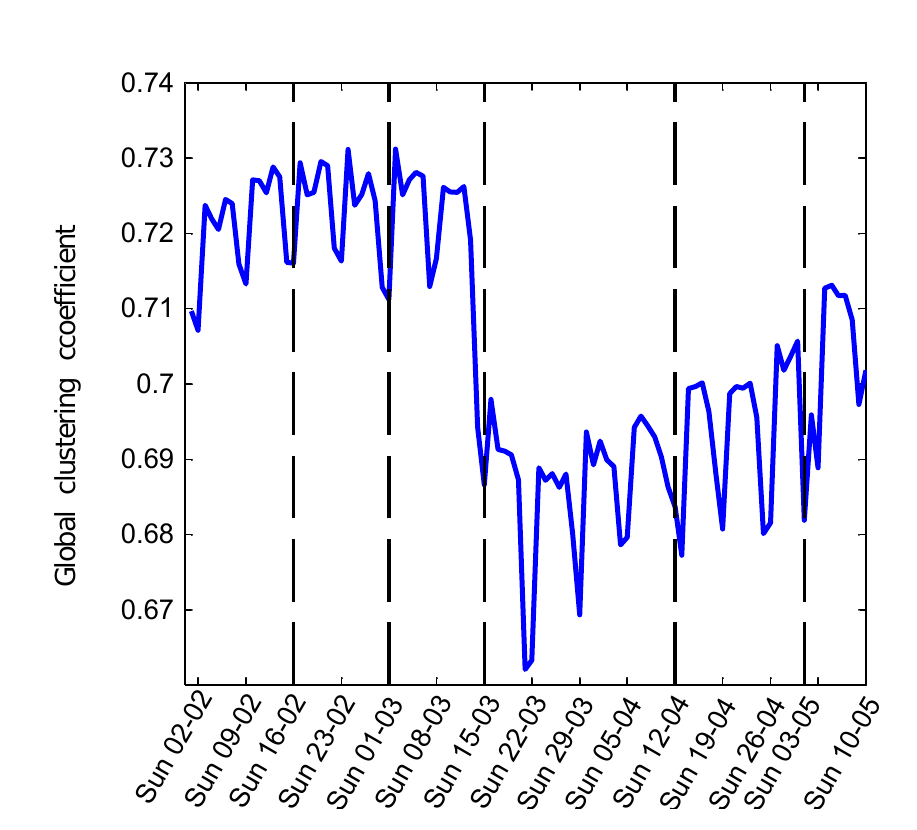}
  \caption{Changes in global clustering value across the Austrian states}
  \label{fig:Communities (Global Clustering)}
\end{figure}

 At first, the global clustering coefficient was calculated for each daily constructed movement graph.
 On the 15\textsuperscript{th} of March (announcement of restrictions) a sharp decrease of this index  can be observed in Figure~\ref{fig:Communities (Global Clustering)}.
 The increase of this metric with a gentle slope after Easter (on 12\textsuperscript{th} of April) is visible in this figure.
 This evidence suggests  a lower connectivity between the nodes caused by the announcement of restrictions. 

\begin{figure}
  \includegraphics[width=\columnwidth]{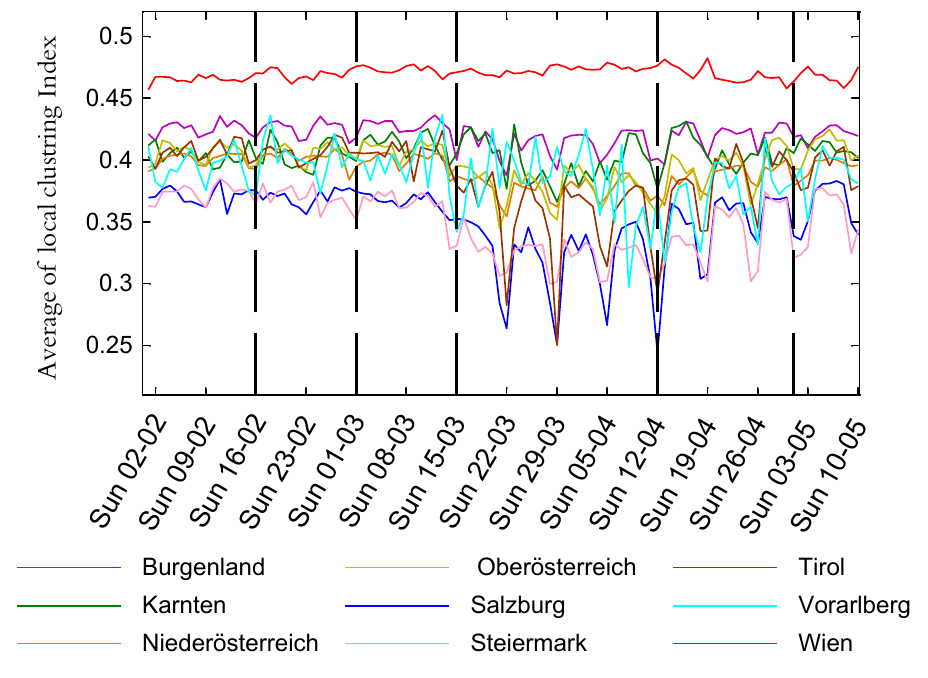}
  \caption{Changes in local clustering values of the Austrian states}
  \label{fig:Communities (Local Clustering)}
\end{figure}

The daily local clustering index was calculated for each political area and then averaged over each federal state.
As is shown in Figure \ref{fig:Communities (Local Clustering)}, each of the states shows a distinctive pattern. For example, the state of Vienna seems to have a steady trend over the all time slots.
On the contrary, states such as Tyrol and Salzburg, where the first positive COVID-19 cases were reported, have the highest reduction rates of local clustering index during the lock-down period.
The reduction of this metric means the interconnection of nodes with neighboring ones are cut down.

At the next step, in order to analyze the structure of crowd movement network at the level of federal states, the aforementioned Newman community detection algorithm \cite{newman2004fast} was utilized.
\begin{figure}
  \includegraphics[width=\columnwidth]{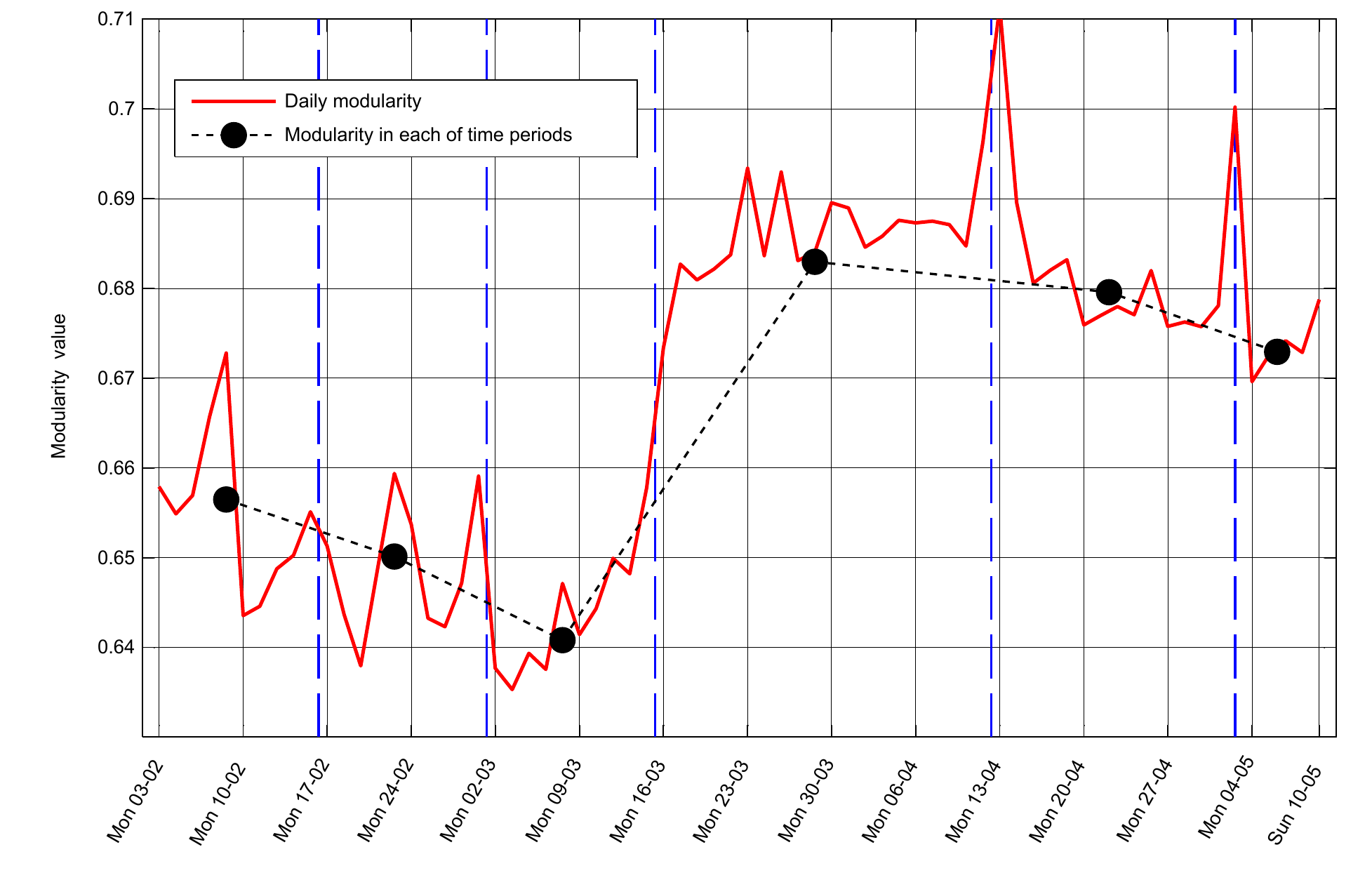}
  \caption{Changes in modularity value of communities}
  \label{fig:Communities (Modularity)}
\end{figure}
Figure \ref{fig:Communities (Modularity)} plots the change of daily modularity value in different time periods.
The modularity value began to increase from announcement of restrictions (on 15\textsuperscript{th} of March) and just after Easter and especially after easing the main restrictions (on 2\textsuperscript{nd} of May) the trend changed (the positive slope has changed to negative). The higher value of modularity in the lock-down period indicates the high concentration of edges in specific groups of nodes and
low interaction among these groups because of the implementation of restrictions.

\begin{figure}
  \includegraphics[width=\columnwidth]{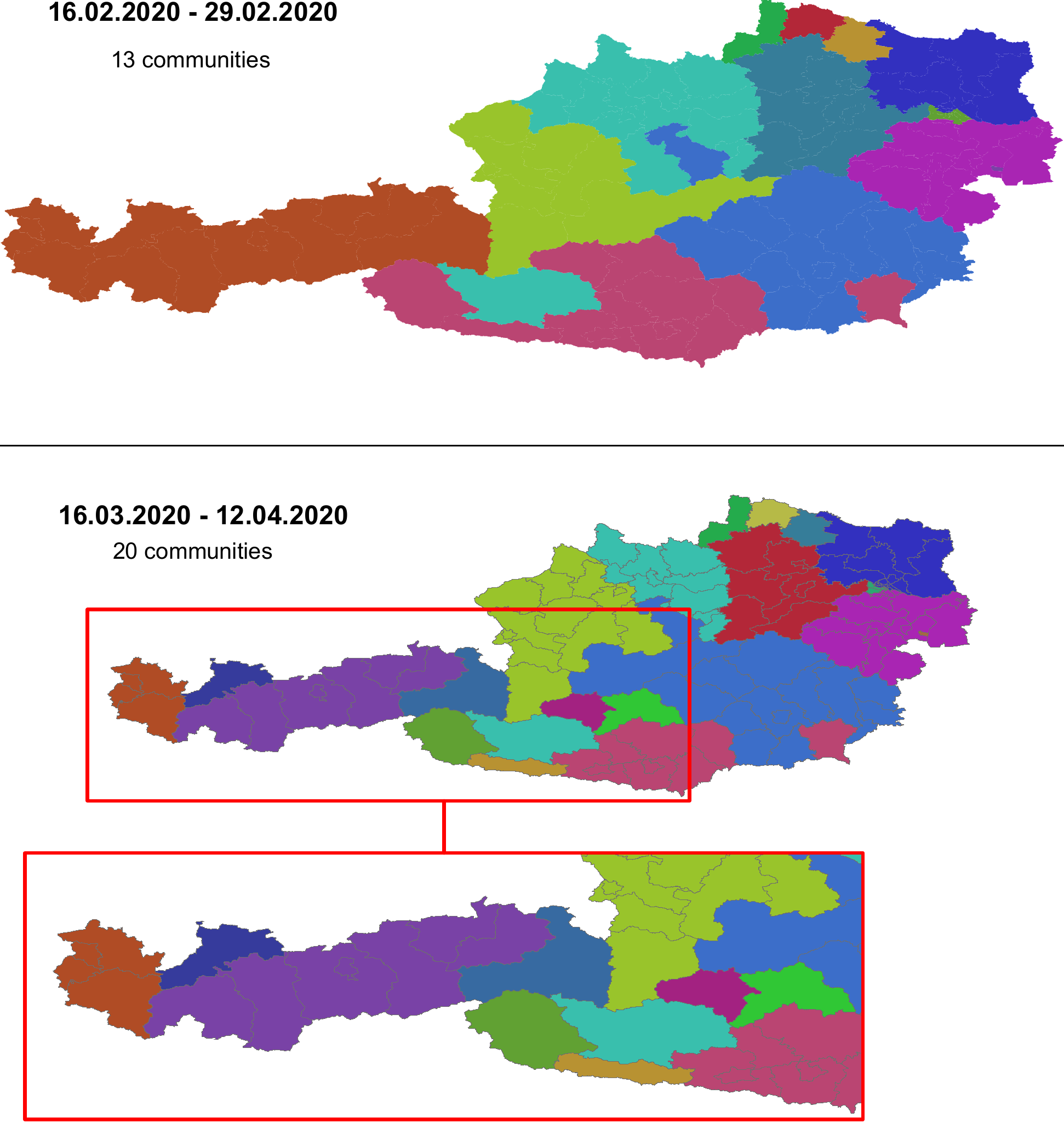}
  \caption{Changes in the communities during the lock-down (most changes happen in the western states, where the pandemic starts)}
  \label{fig:Communities}
\end{figure}

The weighted undirected mobility interaction graphs which were constructed based on the overall \gls{od} matrix in each of defined time periods functioned as an input to the community detection algorithm to yield the best obtained structure of communities in each time slot. The results show the community structure of the mobility interaction network in the form of coherent geographical regions. Figure \ref{fig:Communities} compares the community structure of Austria in two time slots: before initial measures (pre-measures) and after announcement of restrictions.
The count of communities before the lock-down is 13 communities, which increases  during the period of restrictions to 20 communities. As shown in Figure~\ref{fig:Communities}, the subdivision of communities into smaller ones has generally occurred in the western political area of Austria, where the first positive COVID-19 cases were reported and had the most movement reduction rates during the periods of lock-down.
A shrinking size of communities indicates that local interactions are strengthened and mobility to places far away is reduced.

\subsection{Correlation of movement with infection}
\label{h:xischgl}
The first large cluster of infections in Austria originated in Ischgl, a skiing and party area in the Tyrolean alps. In fact, a large share of the infections in Austria, especially at the beginning of the pandemic, could be traced back to this super-spreading event at the beginning of March \cite{ages2020clusterS}.
We can confirm this result by analyzing the \gls{od} matrix constructed in Section \ref{sec:od}.

\begin{figure}
\centerline{\includegraphics[width=\linewidth
]{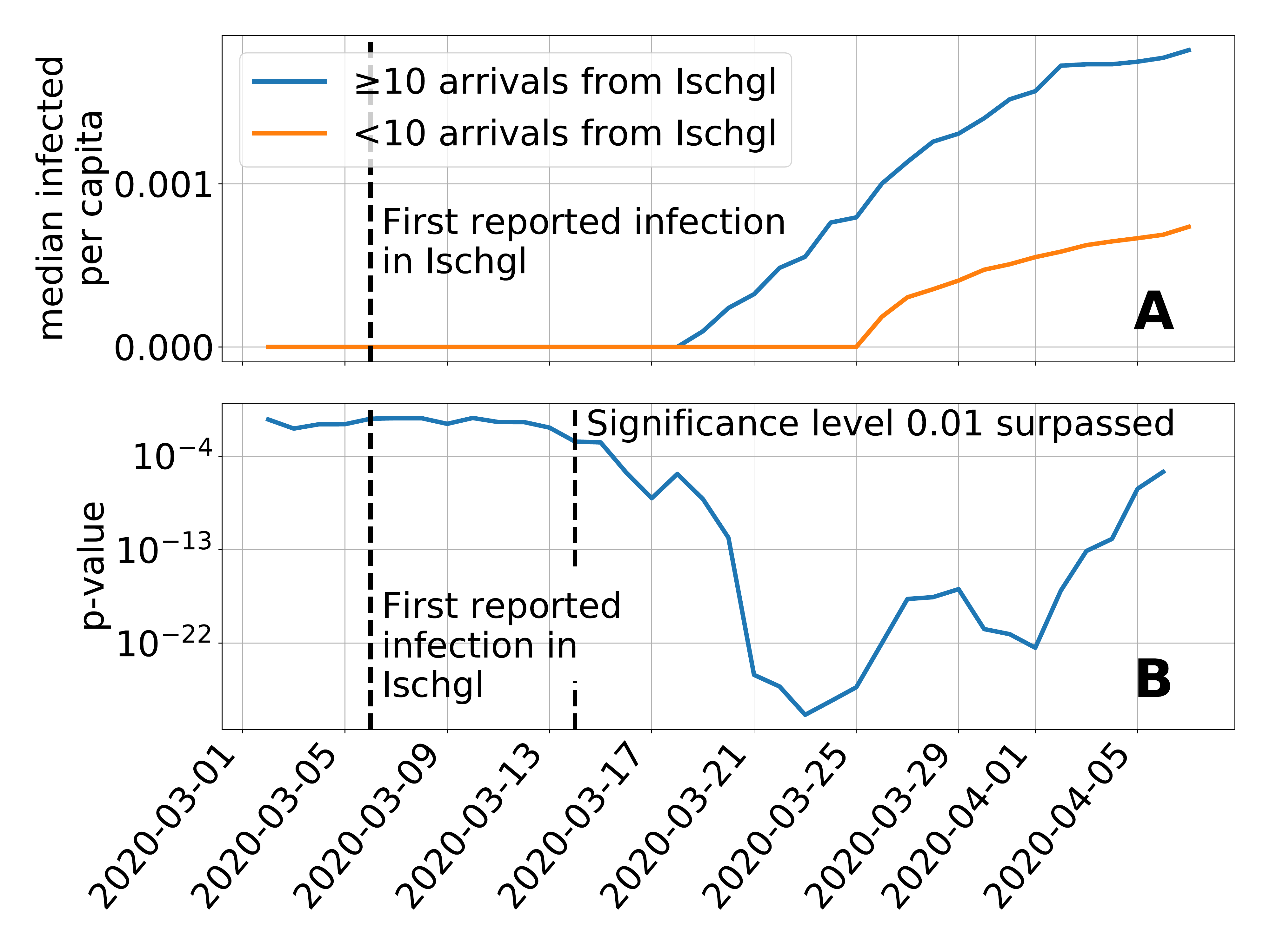}}
\caption{\textbf{A}: Median Infected per capita \textbf{B}: Significance level of a Mann-Whitney test to reject the hypothesis that the two samples are drawn from the same distribution. On March 6\textsuperscript{th} the first cases are reported in Ischgl and on March 14\textsuperscript{th} the effect is significant.
Therefore we observe a time-lag of 8 days with the outflow from Ischgl and the infection rates of the official statistics.)}
\label{fig:infected_ischgl}
\end{figure}

We found that mobile phones leaving the infectious area of Ischgl are spread all over Austria. 
As a next step we split the municipalities into a treatment group with recorded arrivals and a control group with no arrivals from Ischgl.
By comparing the medians of cumulative infection rates of COVID-19 reported by the government (Figure \ref{fig:infected_ischgl}A) we reveal that for towns in the treatment group there is an earlier and stronger onset in COVID-19 infections.

For both groups we compared the relative number of new infections on a municipality level and perform a Mann-Whitney-U significance test.
Figure \ref{fig:infected_ischgl}B shows the resulting p-values starting from March 1\textsuperscript{st}.
We found a significant effect ($p<0.01$) from March 14\textsuperscript{th} onwards, 8 days after the first reported cases in Ischgl (March 6\textsuperscript{th}).

\section{Conclusion}
\ifdraftmode
    \textbf{
    Summary
    }
\fi
In this paper, we described the changes in human mobility in Austria during the lock-down with regards to the SARS-CoV-2 pandemic using near-real-time, anonymized mobile phone data.  We discussed mobility changes for very confined regions such as metro stations, airports or single villages, as well as regional and national changes. The results of statistical, geometric and graph-based methods showed that the announcement of restrictions led to a dramatic reduction in human mobility in the whole country.

\ifdraftmode
    \textbf{
    metric comparison ...
    }
\fi
For all the movement metrics we consistently observed a reduction.
This reduction is followed by an increase in modularity, as communities decompose into clusters.
A similar observation can be made in the other direction as the mobility subsequently recovers.
However, for the \gls{poi} based indicators we cannot observe such a recovery.

\ifdraftmode
    \textbf{
    discussion accuracy.
    }
\fi
Our analyses could be improved if data with a better accuracy were available, such as based on triangulation. 
Other limitations apply as well: we only analyze data from a single big \gls{isp}, its regional market share might vary.
Furthermore, we localize the data only with the coarse cell-id.
As a result, in rural areas the accuracy is in the range of a couple of kilometers, whereas for a city usually approximately 500 m.
In certain rural areas there might be no cell tower coverage at all.
Moreover, people from time to time leave their phone at home, which would not generate movement in our dataset even though people are moving.
    
\ifdraftmode
    \textbf{
    discussion demographics.
    }
\fi

\ifdraftmode
    \textbf{
   discuss ischgl.
    }
\fi
The time lagged correlation of arrivals from Ischgl with infection rates adds to the evidence that origin-destination matrices are of great value for epidemiological forecasting \cite{jia2020population}.
 Due to the correlation we found, we can conclude that between an infection in Ischgl and the detection of the disease in a new area, a duration of maximum 8 days passes.

\ifdraftmode
    \textbf{
   this research demonstrates... 
    }
\fi
We have presented a broad range of measures, some for specific points, others suitable for larger regions.
In total, these offer a global picture of a society with regards to mobility and how the measures have been followed, and can help decision makers by providing input for required simulations and epidemiological models.

\ifdraftmode
    \textbf{
    Future studies...
    }
\fi
A potential future study could analyze co-movement patterns, identify human contacts and subsequently, determine the effect of (non) social distancing \cite{Chen2019} or evaluate the effect of parametrizing the contact behavior from \gls{od} matrices in agent based simulations as employed in e.g. Austria \cite{Bicher2020.05.12.20098970} to predict COVID-cases.
Furthermore, standardizing on a (sub-)set of the presented measures would allow comparison between network providers in the same country or between multiple countries.
This would improve over the analyses of the European commission \cite{JRC2020b, JRC2020, JRC2020a} where inconsistencies between different countries are mentioned.
Additionally, limitations of this study such as the daily re-anonymization of the data could be lifted to perform long-term evaluation of the impact of a pandemic on mobility.

\section*{Acknowledgment}


We are grateful to Eva Bauer for the helpful discussions and support during writing of the paper.


\printglossaries
\printbibliography

\end{document}